\documentclass[preprint,showpacs,double-spaced,aps,prl,floats]{revtex4}
\usepackage{epsfig}
\usepackage{dcolumn}
\usepackage{bm}
\usepackage{amsmath}
\usepackage{amsfonts}
\usepackage{amssymb}
\usepackage{graphicx}%
\newcommand{\bP}{\bar{\Psi}}
\newcommand{\hs}{\hspace*{0.6cm}}
\newcommand{\slA}{\raise.15ex\hbox{$/$}\kern-.57em\hbox{$A$}}
\newcommand{\slB}{\raise.15ex\hbox{$/$}\kern-.57em\hbox{$B$}}
\newcommand{\bA}{\bar{\alpha}}
\newcommand{\bB}{\bar{\beta}}
\newcommand{\beq}{\begin{equation}}
\newcommand{\eeq}{\end{equation}}
\newcommand{\beqn}{\begin{eqnarray}}
\newcommand{\eeqn}{\end{eqnarray}}
\newcommand{\slp}{\raise.15ex\hbox{$/$}\kern-.57em\hbox{$\partial$}}
\newcommand{\pslash}{\raise.15ex\hbox{$/$}\kern-.57em\hbox{$p$}}

\begin{document}
%%%%%%%%%%%%%%%%%%%%%%%%%%%%%%%%%%%%%%%%%%%%%%%%%%%%%%%%%%%%%%%%%%

%%%%%%%%%%%%%%%%%%%%%%%%%%%%%%%%%%%%%%%%%%%%
\title{Critical behavior of the spin correlation function in Ashkin-Teller and Baxter models with a line defect}
%%%%%%%%%%%%%%%%%%%%%%%%%%%%%%%%%%%%%%%%%%%%%%%%%%%%%%%

\author{Carlos Na\'on}
\affiliation{Instituto de F\'\i sica La Plata, CONICET and
Departamento de F\'\i sica, Facultad de Ciencias Exactas,
Universidad Nacional de La Plata, CC 67, 1900 La Plata, Argentina}

\date{April 7, 2009}
%%%%%%%%%%%%%%%%%%%%%%%%%%%%%%%%%%%%%%%%%%%%%%%%%%%%%%%%%%%%%%%%%%%%%%%

\begin{abstract}
We consider the critical spin-spin correlation function of the
Ashkin-Teller and Baxter models. By using path-integral techniques
in the continuum description of these models in terms of fermion
fields, we show that the correlation decays with distance with the
same critical exponent as the Ising model. The procedure is
straightforwardly extended to take into account the presence of a
line defect. Thus we find that in these altered models the critical
index of the magnetic correlation on the defect coincides with the
one of the defective 2D Ising or Bariev's model.
\end{abstract}

\maketitle \hs Two dimensional statistical mechanics systems play a
central role in our present understanding of phase transitions and
critical phenomena. Outstanding members of this family of theories
are the Ising model, the Ashkin-Teller (AT) \cite{AT} and the
eight-vertex (8V) or Baxter models \cite{Bax}. These last two
systems can be mapped onto one another through a duality
transformation. They can be considered as two Ising magnets coupled
by four-spin interactions. They are the first examples of
non-universal critical behavior, in the sense that the critical
exponents of certain operators are continuous functions of the
parameter of the four-spin coupling. An anisotropic version of these
models \cite{anisotropic1}\cite{BaxterBook} leads to an interesting
universality-nonuniversality crossover recently analyzed
\cite{anisotropic2}. Apart from academic interest the AT and 8V
models are useful to shed light on a variety of phenomena, in both
classical and quantum physics, ranging from biological applications
\cite{DNA} to the theory of cuprate superconductors \cite{cuprate}.

Concerning the isotropic case, which we will consider in this work,
it was conjectured that the magnetization keeps the Ising behavior,
with a universal exponent $\Delta_{\sigma}=1/8$
\cite{conjecture-Baxter} \cite{conjecture-AT}. This result was later
proved by Baxter through corner transfer matrices \cite{BaxterBook}.
It is however surprising that there is no other direct computation
of the two-spin correlation function in the literature.

Much less is known exactly about the behavior of these systems in
the presence of line defects \cite{Igloi-Peschel-Turban}. For the
simpler Ising lattice with an altered row (Bariev's model
\cite{Bariev}) it has been shown that the scaling index of the
magnetization varies continuously with the defect strength
\cite{Bariev, McCoy-Perk}, whereas the critical exponent of the
energy density at the defect line remains unchanged \cite{Brown,
Ko-Yang-Perk, Burkhardt-Choi}. Taking this model as working bench,
much insight was obtained about the origin of nonuniversal critical
behavior. For instance, in Ref.
[\onlinecite{Burkhardt-Eisenriegler}] necessary conditions for the
dependence of exponents on the coupling constants were derived.
Interesting connections with integrable quantum field theories were
also revealed \cite{Mussardo}.

Despite these important advances the behavior of the spin-spin
correlator for critical 8V-AT models with line defects remains
unknown. The main goal of this paper is to help filling this gap. We
shall derive a central feature of that critical behavior through a
straightforward calculation performed within the continuous
formulation of AT-8V models, using well established path-integral
techniques. Since AT-8V models have both magnetic and electric
correlations \cite{BaxterBook} (the electric correlations have
continuously varying exponents \cite{Ko-electric}), we stress that
in this paper we will be concerned with magnetic correlations only.
We will show that the magnetic exponent depends on the strength of
the defect in exactly the same way as in Bariev's model. In this
way, our result provides a very unusual explicit confirmation of
universality.

Since it is crucial for our purpose to use a path-integral approach
that allows to factorize the Ising correlator, and this method works
for both the usual and altered cases, in order to illustrate it, we
start by considering the first case corresponding to the homogeneous
lattice. This intermediate step will provide an analytical argument
for the behavior of the two-spin correlation function
\cite{conjecture-Baxter}\cite{conjecture-AT}. At the end of the
paper we will show how the main result, valid for the altered
models, is obtained.

The Hamiltonian of the original lattice model is given by
\begin{equation}
{\cal H} = -\sum_{<ij>} \big(J_2 \, (\sigma_i \sigma_j + \tau_i
\tau_j) + J_4 \,\sigma_i \sigma_j \tau_i \tau_j \big)
\end{equation}
where $<ij>$ means that the sum runs over nearest neighbors of a
square lattice ($\sigma,\tau=\pm 1$).

As shown in Ref. [\onlinecite{LP}] the scaling regime of AT-8V
models can be described in the continuum limit in terms of a
Thirring-Luttinger Lagrangian, i.e. a model of Dirac fermions
coupled by a quartic interaction. Alternatively, this can be
expressed as two Majorana fermions interacting via their
energy-densities:
\begin{equation}\label{lagrangian}
{\cal L}[\alpha, \beta] = \bA i\slp \alpha + \bB i\slp \beta -
\lambda \, \epsilon_{\alpha}\,\epsilon_{\beta}
\end{equation}
where $\alpha$ and $\beta$ are the Majorana spinors with components
$\alpha_{1,2}$, $\beta_{1,2}$ respectively. Let us recall that this
components are connected to fermion annihilation and creation
operators $c_r$ ($d_r$) and $c^{\dagger}_r$ ($d^{\dagger}_r$)
attached to site $r$ ($c_r =
\frac{e^{-i\pi/4}}{\sqrt{2}}(\alpha_{1}(r)+i\alpha_{2}(r)$), $d_r =
\frac{e^{-i\pi/4}}{\sqrt{2}}(\beta_{1}(r)+i\beta_{2}(r)$)).
$\epsilon_{\alpha}=\alpha_{1} \alpha_{2}$ and
$\epsilon_{\beta}=\beta_{1} \beta_{2}$ are the corresponding
energy-densities. The symbol $\slp$ stands for $\gamma_{\mu}
\partial_{\mu}$, with $\gamma_{\mu}$ the usual Euclidean Dirac matrices ($\mu=0,1$ associated to space directions).
The coupling constant $\lambda$ is proportional to $J_4/J_2$.

Similar manipulations, based on the Jordan-Wigner transformation
\cite{SML}, allow to write the on-line spin-spin correlation
function in the form \cite{BI}
\begin{equation}\label{correlation}
<\sigma(0)\sigma(R)> = <\exp\;({\pi \int_{0}^{R}dx\,
\epsilon_{\alpha}(x)})>
\end{equation}
where the vacuum expectation value is an anticommuting path-integral
to be evaluated with the continuum action $S = \int d^2x\,{\cal L}$,
with an integration measure $\cal{D}\alpha\,\cal{D}\beta$. For
$\lambda =0$ the $\beta$-fields become completely decoupled and the
computation can be readily performed either in terms of the Majorana
$\alpha$-fields or in terms of Dirac fermions \cite{ZI} built
through the doubling technique \cite{Ferrell}, yielding the
well-known result for the Ising correlator. In what follows we will
show how this last procedure can be extended in order to allow for a
tractable route leading to the exact computation of the critical
exponent of the spin-spin correlation function for 8V and AT models.
We start by squaring (\ref{correlation}):
\begin{equation}\label{squared}
<\sigma(0)\sigma(R)>^2 = <\exp\;\big({ \pi \int_{0}^{R}dx\,
(\epsilon_{\alpha}(x)+\epsilon_{\alpha'}(x))}\big)>
\end{equation}
where the vacuum expectation value must now be computed with respect
to an Euclidean action with Lagrangian density $\tilde {\cal
L}[\alpha,\beta,\alpha',\beta']= {\cal L}[\alpha,\beta]+ {\cal
L}[\alpha',\beta']$, $\alpha'$ and $\beta'$ being the replicated
fermion fields. Following Ref.\onlinecite{ZI} we can build Dirac
fermions $\Psi$ and $\chi$ as the following combinations:
\begin{equation}
\Psi= \alpha + i \alpha', \,\,\,\chi= \beta + i \beta'.
\end{equation}

In terms of these new fields we can write the Lagrangian $\tilde
{\cal L}[\alpha,\beta,\alpha',\beta']$ in the form
\beqn\label{newLagrangian}&\tilde {\cal L}[\Psi,\chi]&=\bP i\slp
\Psi + {\bar{\chi}} i\slp \chi - \frac{\lambda}{8}
\,\Big({\bar{\chi}}\,\gamma_5 \, \chi \,\bP \, \gamma_5 \, \Psi +
Im(\chi^T \gamma_1 \chi^T)Im(\Psi^T \gamma_1 \Psi^T)\Big), \eeqn
where $\gamma_5 =i\gamma_0 \gamma_1$, and $\Psi^T,\chi^T$ are the
transposed spinors. On the other hand equation (\ref{squared}) can
be expressed as
\begin{equation}\label{final-squared}
<\sigma(0)\sigma(R)>^2 = <\exp\;\Big({\pi \int
d^2x\,\bP\,\slA\,\Psi}\Big)>,
\end{equation}
where now the path integral integration measure in the right hand
side is expressed in terms of the fields $\Psi$ and $\chi$, and
$A_{\mu}$ is an auxiliary vector field with components: \beq
\label{Amu1} A_{0}(x_{0},x_{1}) = \delta(x_{0}) \theta(x_{1})
\theta(R-x_{1}), \,\,A_{1}(x_{0},x_{1}) = 0 . \eeq At this point we
note that a similar manipulation has been earlier introduced
\cite{Na} and employed to compute several correlation functions in
both Ising and 8V models \cite{Fer-Na}. In those cases this method
allowed to identify the objects to be computed with certain
fermionic determinants that could be evaluated through an
appropriate change of path-integral variables. At first sight one
sees that the problem is much more involved in the present case,
since the correlation is not directly associated to a simple
fermionic determinant. Indeed, gathering the above results we can
write:
\begin{equation}\label{partitions}
<\sigma(0)\sigma(R)>^2 = \frac{Z[g=\pi]}{Z[g=0]},
\end{equation}
where \begin{equation} Z[g]=\int{\cal D}\bP{\cal D}\Psi{\cal
D}{\bar{\chi}}{\cal D}\chi\exp\big({-\int d^{2}x\big(\tilde {\cal
L}[\Psi,\chi]- g \bP \slA\,\Psi \big)}\big).\end{equation}

This is our first non-trivial result. The continuum limit of the
squared two-point spin correlation function is {\em exactly}
expressed in terms of the vacuum to vacuum functional of a quantum
field theory describing two interacting fermion species. We now show
how the right hand side of (\ref{partitions}) can be put as the
product of two $R$-dependent factors, one of which being the squared
spin-spin correlator of the Ising model. To this end we make the
following change of path-integral variables in the numerator of
equation (\ref{partitions}), with chiral and gauge parameters $\Phi$
and $\eta$, respectively: \beq \label{changeofvar1} \Psi =
e^{-\pi(\gamma_{5}\Phi+i\eta)}\;\zeta,\,\, \bP = {\bar{\zeta}}\;
e^{-\pi(\gamma_{5}\Phi-i\eta)}. \eeq Note that the $\chi$-fields are
left unchanged. If the parameters of the transformation are related
to the previously introduced vector field $A_{\mu}$ in the form \beq
\label{Amu}A_{\mu} = \epsilon_{\mu\nu}\partial_{\nu}\Phi +
\partial_{\mu}\eta \eeq
then the only $R$-dependent term in the action (i.e. the one with
"coupling constant" $g=\pi$) becomes completely decoupled. As the
result of the change the dependence on $R$ reappears in two places:
in the $\lambda$-term that couples both fermion species $\chi$ and
$\zeta$, through the dependence of $\Phi$ and $\eta$ on $R$, and in
the Jacobian associated to (\ref{changeofvar1}).
 As explained in Ref. [\onlinecite{Nao}], this Jacobian must be computed with
a gauge-invariant regularization prescription in order to avoid an
unphysical linear divergence. Following this procedure it has been
shown that the Jacobian exactly coincides with the squared critical
spin-spin function of the Ising model \cite{Na}. Then we have
arrived at the following identity:

\begin{equation}
<\sigma(0)\sigma(R)>^2 = <\sigma(0)\sigma(R)>_{Ising}^2
\,F(\lambda,R),
\end{equation}
with \beq \label{F}F(\lambda,R) = {\cal N}(\lambda)\,
<\exp{\big(S_{\Phi}(\zeta,\chi)+S_{\eta}(\zeta,\chi)\big)}>_0,\eeq
where $<>_0$ means vacuum expectation value with respect to the
model of free $\chi$ and $\zeta$ fermions. ${\cal N}(\lambda)$ is a
normalization constant independent of $R$. Since the analysis of the
dependence of $F(\lambda,R)$ on $R$ is more easily done in momentum
space we have Fourier-transformed $S_{\Phi}(\zeta,\chi)$ and
$S_{\eta}(\zeta,\chi)$ in the above equation: \beqn &
&S_{\Phi}(\zeta,\chi)=\frac{\lambda}{8}\int \prod_{j=1}^4
\frac{d^{2}p_j}{(2\pi)^2} \, {\bar{\chi}}(p_1)\,\gamma_5 \,
\chi(p_2) \,{\bar{\zeta}}(p_3) \, \gamma_5 \,G(P) \zeta(p_4), \eeqn
with $G(P,R)$ a diagonal $2\times 2$ matrix given by
\begin{equation}\label{G}
G(P,R)=
\begin{pmatrix}
g_+(P,R) & 0 \\
0 & g_-(P,R)
\end{pmatrix}
\end{equation}
where $g_{\pm}(P,R) = \pm \int d^{2}x\,e^{i P\cdot x}e^{\mp
2\pi\,\Phi(x,R)}$, and $P=p_1 + p_2 + p_3 +p_4$. A similar
expression is obtained for $S_{\eta}$ with $G(P,R)$ replaced by
\begin{equation}\label{H}
H(P,R)=
\begin{pmatrix}
h(P,R) & 0 \\
0 & h(P,R)
\end{pmatrix}
\end{equation}
with $h(P,R)=\int d^{2}x\,e^{i P\cdot x}e^{2i\pi\,\eta(x,R)}$.

The explicit functional forms of $\Phi(x,R)$ and $\eta(x,R)$ can be
determined by combining equations (\ref{Amu1}) and (\ref{Amu}),
which yields $\Phi(x,R)=-\frac{1}{4\pi}\,\log\frac{x_0^2 + (R-x_1)^2
+ a^2}{x_0^2 + x_1^2 + a^2}$, and
$\eta(x,R)=\frac{x_0}{2\pi}\,\int_0^R \frac{dy}{x_0^2 + (y-x_1)^2 +
a^2}$, where $a$ is an ultraviolet cutoff which can be identified
with the lattice spacing of the original discrete system.

Our problem is now reduced to the analysis of the integrals
$g_{\pm}(P,R)$ and $h(P,R)$. In so doing one notes that the
integrals diverge for large distances, which leads us to the
introduction of a cutoff $L$ which can be interpreted as the size of
the system (of course, the thermodynamic limit will be recovered by
setting $L \rightarrow \infty$ at the end of the computation). In
terms of the dimensionless variables $u_\rho =\frac{x_\rho}{L}$ (
$\rho=0,1$), we obtain \beqn\label{g} & &g_{\pm}(P,R) = \lim_{L
\rightarrow \infty}\pm L^2 \int_{\mid u_\mu \mid < 1} d^{2}u\, e^{i
L\,P\cdot u} \, \Big(\frac{u_0^2 + (u_1-(R/L))^2 + a^2/L^2}{u_0^2 +
u_1^2 + a^2/L^2}\Big)^{1/2}=\nonumber\\ &=&\pm (2\pi)^2 \delta^2(P)
\eeqn and a similar result for $h(P,R)$. Then, it is apparent now
that in the thermodynamic limit ($a \ll R\ll L$) $F(\lambda,R)$
becomes independent of $R$ and the critical behavior exactly
coincides with the one of the $2D$ Ising model. This provides an
analytical argument for the conjectures first given in Refs.
[\onlinecite{conjecture-Baxter}] and [\onlinecite{conjecture-AT}].

Let us now address the main issue of this work. We include a line
defect in one of the original Ising lattices, say the one with spins
$\sigma$. To be specific we consider the so called chain defect
(here we employ the terminology of Ref.
[\onlinecite{Igloi-Peschel-Turban}], which corresponds to Bariev's
second type defect, in which bonds along the same column are
replaced: $J_2 \rightarrow J_2'$). We will study the two-spin
correlation function in the column of altered bonds ($x_0=0$)
\cite{McCoy-Perk}. In passing we recall that this model could be
mapped on to an XYZ spin $1/2$ quantum chain \cite{BaxterBook}. In
this framework, seeing $x_0$ as the temporal variable associated to
the quantum evolution, our defect maps on to an additional
interaction affecting the whole chain (in contrast to a line defect
parallel to the time axis that would map on to an individual
impurity site). It is known that the continuous version of the
classical model is modified, due to the defect, by the addition in
equation (\ref{lagrangian}) of a term
$2\pi\mu\,\delta(x_0)\,\epsilon_{\alpha}(x)$, with $\mu=J_2'-J_2$.
By carefully examining the fermionic representation of $\sigma$-spin
operators on the lattice, following the lines of Ref. \cite{ZI}, one
also finds that in the continuum limit each spin operator on the
defect line picks up a similar $\mu$-dependent factor, in such a way
that the squared correlator for the defective model is given by a
simple modification of equations (\ref{final-squared}) and
(\ref{Amu1}):
\begin{equation}\label{string}
<\sigma(0)\sigma(R)>_{\mu}^2 = <\exp\Big({\pi (1+4\mu)\int
d^2x\,\bP\,\slA\,\Psi}\Big)>_{\mu}.
\end{equation}
At this point we stress that this formula is valid for
$\sigma$-spins. The $\mu$-dependence of the coefficient comes from
the fact that the additional defect term mentioned above can be
interpreted as a position-dependent mass term for $\Psi$-fermions.
This additional mass term is not present for $\chi$-fermions, and
consequently, the corresponding coefficient in the string
representation for $<\tau(0)\tau(R)>_{\mu}^2$ will be just equal to
$\pi$.
 From now on one can follow exactly the
same steps already explained for the defect-free case. The presence
of the defect manifests in the exponent of the path-integral change
of variables given by equations (\ref{changeofvar1}), where one has
to make the substitution $\pi \rightarrow \pi(1+4\mu)$. Once again
the squared two-point function factorizes in the form
\begin{equation}\label{squared-defective}
<\sigma(0)\sigma(R)>_{\mu}^2 = <\sigma(0)\sigma(R)>_{Bariev}^2
\,F(\lambda, \mu, R).
\end{equation}
For the first factor in the right hand side of this equation we
obtain, through the corresponding Jacobian, the well-known behavior
for Bariev's model: $<\sigma(0)\sigma(R)>_{Bariev}\simeq
(\frac{a}{R})^{2\Delta_{\sigma}}$, with
$\Delta_{\sigma}=\frac{1}{8}(1+4\mu)^2$
[\onlinecite{Bariev,McCoy-Perk,Cabra-Naon}]. Concerning the second
factor, $F(\lambda, \mu, R)$ has the same structure as $F(\lambda,
R)$, already depicted in equations (\ref{F}),(\ref{G}) and (\ref{H})
(in fact it satisfies $F(\lambda, 0, R)=F(\lambda, R)$). The only
difference is the appearance of the $\mu$-dependent factors in the
exponents of the integrals defining the functions $g_{\pm}(P,R)$ and
$h(P,R)$. Therefore, it turns out that these integrals can be
written following the same prescription as in the $\mu=0$ case. The
corresponding expression coincides with (\ref{g}), with an exponent
$\kappa = \frac{1}{2}(1+4\mu)$ instead of $\frac{1}{2}$ in the
fraction of functions. A similar result is obtained for $h(P,R)$.
Let us mention that the constant $\kappa$ satisfies $\mid \kappa
\mid <1$. Indeed, as shown in Ref. [\onlinecite{Burkhardt-Choi}],
the allowed range of defect strengths in the lattice Ising model
corresponds to $\mid \mu \mid < \frac{1}{4}$ in its continuous
version. Since for $\lambda =0$ we should reobtain the results
corresponding to Bariev's model, we conclude that for the present
models the above condition also coincides with the physically
relevant interval of defect strengths. But the main point is that in
the thermodynamic limit ($a \ll R\ll L$), we find that $F(\lambda,
\mu; R)$ becomes independent of $R$, and the critical exponent for
the magnetic correlation is unchanged by the coupling between the
different Ising subsystems, keeping the same value as in the 2D
Ising model with a defect line. This is our main result. It is worth
noting that the irrelevance of the $\lambda$-coupling is consistent
with a scaling picture, similar to the one presented in
\cite{Burkhardt-Eisenriegler}. Indeed, if one sees the coupling on
the defect as a perturbation, its scaling dimension is $d - 1
-2\Delta_{\epsilon}$, where
$\Delta_{\epsilon}=\frac{1}{1+2\lambda/\pi}$ is the dimension of the
energy-density. Since $d=2$ one finds that the coupling is
irrelevant for $\mid\lambda\mid<\pi/2$.

Concerning the correlations between $\tau$-spins, the computation is
more subtle. Since $\tau$-spins are related to $\chi$-fermions, one
has to perform the transformation (\ref{changeofvar1}) not only for
the $\zeta$-fields, but for the $\chi$-fields also. However, due to
the fact that the defect only affects the couplings between
$\sigma$-spins, the string representation for $\tau$-correlations on
the defect line, $<\tau(0)\tau(R)>_{\mu}^2$ does not pick up any
$\mu$-dependent coefficient, in contrast to what happened for
$<\sigma(0)\sigma(R)>_{\mu}^2$ (see equation (\ref{string})). One
then finds that these correlations decay as in the homogeneous Ising
model.

To summarize, we have determined the critical behavior of the
two-spin correlation in the continuum, field-theory version of
isotropic AT and 8V models. This scheme treats in a unified way both
the homogeneous (defect-free) and inhomogeneous (with a line
defect). In the first case we provided an analytical derivation for
the value of the magnetic exponent, $\Delta_{\sigma}=\frac{1}{8}$.
In the second case we found that on the line defect the critical
index maintains Bariev's value:
$\Delta_{\sigma}=\frac{1}{8}(1+4\mu)^2$, where $\mu$ is the strength
of the defect. The critical index corresponding to the spins with
homogeneous couplings remains equal to the Ising value in all cases
($\Delta_{\tau}=\frac{1}{8}$). Our result is an explicit
confirmation of universality, in the sense that the interactions
between the elementary Ising subsystems do not affect the critical
exponents, even in the presence of a line defect. Besides these new
results, we hope that our approach will be useful to shed light on
interesting problems concerning inhomogeneous AT-8V models. In
particular it could be used to analyze scaling properties at
interfaces between critical subsystems \cite{LTI}.

\vspace{1cm}

\noindent {\bf Acknowledgement}\\
The author is grateful to CONICET and UNLP (Argentina) for financial
support. He thanks D. Cabra, G. Rossini and M. Salvay for useful
discussions and A. Iucci for helpful correspondence.

\end{document}